\documentclass[12pt]{article}
\usepackage{epsfig}
\def\be{\begin{equation}}
\def\ee{\end{equation}}

\def\lsim{\raise0.3ex\hbox{$<$\kern-0.75em\raise-1.1ex\hbox{$\sim$}}}
\def\gsim{\raise0.3ex\hbox{$>$\kern-0.75em\raise-1.1ex\hbox{$\sim$}}}

\def\NP{{ Nucl.\ Phys.\ }}

\begin{document}

\vskip 1.5 cm

\centerline{\large{\bf Effective Z(2) Spin Models of}}

\medskip

\centerline{\large{\bf Deconfinement and Percolation in SU(2) Gauge Theory}}

\vskip 1.0cm

\centerline{\bf S. Fortunato, F. Karsch, P. Petreczky and H. Satz}

\bigskip

\centerline{Fakult\"at f\"ur Physik, Universit\"at Bielefeld}
\par
\centerline{D-33501 Bielefeld, Germany}

\vskip 1.0cm

\noindent

\centerline{\bf Abstract:}

\medskip

Effective theories are helpful tools to
gain an intuitive insight into phenomena  governed by
complex laws. In this work we 
show by means of Monte Carlo simulations 
that  $Z(2)$ spin models with only spin-spin interactions
approximate rather well the critical behavior of $SU(2)$
gauge theory at finite temperature. 
These models are then used to formulate an effective description of
Polyakov loop percolation, which is shown to reproduce the thermal critical
behaviour of $SU(2)$ gauge theory.
\vskip 1cm

\section{Introduction}

The global $Z(N)$ center symmetry plays a key role
in the confinement-deconfinement transition of 
$SU(N)$ gauge theories.  
The spontaneous breaking of this symmetry is  responsible
for the phase transition, and the corresponding order parameter is
the Polyakov loop \cite{mclerran81,svetitsky82}. In full QCD
the dynamical quarks act as a (small) external magnetic field \cite{green84}
and the conventional formalism of spontaneous symmetry breaking
does not apply. Therefore it seems  helpful to consider
an alternative approach to critical behavior of $SU(N)$
gauge theory, which may be more readily generalizable to full QCD.

In \cite{satz98} it was suggested that the deconfinement
phase transition in $SU(N)$ gauge theory can be viewed as a percolation 
transition of suitably defined Polyakov loop clusters. 
In \cite{fortunato00,aci} such an approach 
has been applied to (2+1)- and (3+1)-dimensional $SU(2)$ pure gauge theory;
the clusters are formed by joining nearest-neighbour like-sign Polyakov
loops with a bond probability similar 
to the one used to define the ``physical droplets'' 
in the Ising model \cite{coniglio80}.
In \cite{fortunato00,aci}, the bond probability
was determined from an effective theory of Polyakov loops,
which was derived by a strong coupling expansion \cite{green84}.
Hence the validity of this approach 
is limited to the strong coupling limit of $SU(2)$, and
the calculations were thus done on lattices with a temporal extent of only
two lattice spacings.
It is thus not clear how to formulate a percolation picture 
for $SU(2)$ in the continuum limit.

The aim of the present paper is to clarify whether the
deconfinement phase transition in (3+1)-dimensional $SU(2)$ gauge theory 
can be described more generally as a percolation transition. 
We will derive an effective
theory for $Z(2)$ spin variables containing up to 19 spin-spin
couplings which are obtained by solving Schwinger-Dyson equations, as 
originally proposed in \cite{gonzales87,okawa88}.
This method is independent of the number $N_{\tau}$ of lattice spacings 
in the time direction and we will apply it to two different cases, 
$N_{\tau}=2$ and $4$. The particular 
effective theory studied here is then shown to allow an equivalent percolation 
formulation of the deconfinement transition.
The corresponding
clusters are, with good approximation, 
the "physical droplets" of $SU(2)$.

\section{Percolation and the Effective Theory}

We are interested in constructing an effective theory
of Z(2) spin variables $\{s_{\bf n}=\pm 1\}$,
defined as the signs of  Polyakov loops at the spatial sites ${\bf n}$.
The effective Hamiltonian
${\cal H}(s,\beta)$ of the signs $\{s_{\bf n}\}$ 
of the Polyakov loop configurations can be defined through 
the equation \cite{gonzales87,okawa88}
\begin{equation}
\label{effth}
\exp[{\cal H}(s,\beta)]\,=\,\int\,[dU]\prod\limits_{\bf n}\,\delta[s_{\bf n}, sgn(L_{\bf n})]
\,\exp(-S_{W}(\beta)),
\end{equation}
where $L_{\bf n}$ is the 
value of the Polyakov loop at the spatial point {\bf n} and 
$S_{W}$ the Wilson lattice action which depends on the gauge coupling $\beta$. 
Eq. (\ref{effth}) shows that 
all degrees of freedom of the 
original $SU(2)$ field configurations
are integrated out, leaving only the 
distribution of the corresponding
Ising-projected configuration.

The problem is now how to determine 
the explicit form ${\cal H}(s,\beta)$, starting from the 
original Polyakov loop configurations.
In general, ${\cal H}(s,\beta)$ will contain infinitely many operators.
However, in the vicinity of the critical point where the correlations
become long ranged one expects that
${\cal H}(s,\beta)$ can be well approximated by a finite sum of different
terms \cite{svetitsky82}. So we write
\begin{equation}
\label{effham}
{\cal H}(s,\beta)\,=\,\sum_{\bf n}\sum_{\alpha}
\kappa_{\alpha}(\beta)O^{\alpha}_{\bf n},
\end{equation}
where $\kappa_{\alpha}(\beta)$ are the couplings,
$O^{\alpha}_{\bf n}$ are the spin-spin operators at site ${\bf n}$ and
$\alpha$ specifies the specific spin-spin interaction;
e.g., in the simplest case
there would be only a single term 
$\kappa_1 O_{\bf n}^1 = 
\kappa_1 \sum_{\hat \mu=1}^4 s_{\bf n} s_{{\bf n}+ \hat \mu}$,
with $\hat \mu$ denoting a unit vector in $\mu$-direction.
In contrast to  \cite{okawa88}, we here do not include multi-spin operators
in Eq. (\ref{effham}), because it is not known how to
define clusters in this case.

To calculate the couplings, 
we use the following 
set of Schwinger-Dyson equations
\begin{equation}
\label{dyson}
\langle O^{\gamma}_{\bf n} \rangle\,=
\,-\langle O^{\gamma}_{\bf n}\,
\exp(-2 \sum_{\alpha}\kappa_{\alpha}O^{\alpha}_{\bf n}) \rangle,
\end{equation}
which are derived by exploiting the $Z(2)$
symmetry of ${\cal H}(s,\beta)$ \cite{gonzales87}.
Here $<...>$ denotes   averages taken in the $(3+1)$-dimensional
$SU(2)$ gauge theory.
Eqs. (\ref{dyson}) establish a relation
between thermal averages of the operators $O^{\gamma}_{\bf n}$ and
the couplings $\kappa_{\gamma}$.
These equations are, however, implicit in the couplings.
They can be solved by means of the
Newton method, based on 
successive approximations \cite{okawa88}.
One starts by 
making a guess about the values of the couplings and derives
from Eqs. (\ref{dyson}) another set of values for $\kappa_{\gamma}$'s.
After 
a sufficient number of iterations,
the series of partial values for the $\kappa_{\gamma}$'s will converge
to the solution of Eqs. (\ref{dyson}).

We notice that the general set of equations
(\ref{dyson}) refers to a single point 
{\bf n} of the spatial volume. 
However, the thermal averages are independent of 
the particular point {\bf n}, so it doesn't matter
where we decide to take the averages.
We thus explore translational invariance and the Eqs. (3) should thus
be understood as being averaged over the entire lattice.
This reduces considerably the effect 
of thermal fluctuations  
and, consequently,
the errors on the final $\kappa_{\gamma}$'s.

With this we have  all necessary tools to derive
an effective theory for $SU(2)$ from the Polyakov loop configurations.

In general, the approximation improves the more operators
we include in (\ref{effham}).
The fact that one must restrict
the choice to some subset of operators involves
a truncation error in addition to the statistical one.
The error due to truncation is difficult to control.
It can, however, be estimated by calculating the parameters of the effective theory 
for different sets of operators.
We also need  to establish a criterium to judge how
well the effective theory approximates the original one. 
A good option 
could be to compare average values we get from 
the configurations produced by simulating the effective theory with the 
corresponding quantities measured on the
original Ising-projected 
Polyakov loop configurations. In particular, we used the lattice average of the
magnetization $m$, 
\begin{equation}
\label{magneff}
m\,=\,\frac{1}{V}|\sum_{i}s_i|,
\end{equation}
($V$ is the spatial lattice volume) for such a test.

We point out that the approach we have described
can be used for any value of the number
$N_{\tau}$ of 
lattice spacings in the temperature
direction, although 
we have applied the method here only to $SU(2)$ in $(3+1)$ dimensions for
$N_{\tau}=2$ and $4$. 

The magnetization transition in the Ising model
can be described as percolation transition of appropriately
defined site-bond clusters \cite{coniglio80}. The clusters
are defined by the condition that two nearest neighboring spins
of the same sign belong to a cluster with {\em bond probability}
(or bond weight)
\begin{equation}
p=1-\exp(-2 \kappa). 
\label{io}
\end{equation}
Here $\kappa$ is the Ising coupling divided by the temperature.

It was recently  proved that 
this result can be extended to 
other spin models \cite{san1,san2,san3}.
In particular, if 
a theory contains only ferromagnetic 
spin-spin interactions, the above percolation picture can be trivially
extended by introducing a bond 
between each pair of interacting spins,
and a relative bond probability
\begin{equation}
\label{tu}
p_{\alpha}=1-\exp(-2 \kappa_{\alpha}), 
\end{equation}
where $\kappa_{\alpha}$ is the coupling 
associated to the $\alpha$-th interaction of the theory.

As we will see in the next section, all couplings
of the effective theory indeed turn out to be  positive.  
To  carry out the percolation studies we will 
build the clusters in the Polyakov loop configurations using bond
probabilities given by Eq. (\ref{tu}), with $\kappa_{\alpha}$ values taken
from the effective theory. 
Once we have grouped all spins into clusters, we calculate 
the values of the following
variables \cite{stauffer94}:
\begin{itemize}
\item The {\it percolation strength } $P$, 
      defined as the probability that a randomly chosen lattice site 
      belongs to the percolating cluster. 
      $P$ is the {\it order parameter} of the percolation transition.

\item The {\it average cluster
        size } $S$, defined as 
      \begin{equation}
        S=\frac{\sum_{s} {{n_{s}s^2}}}{\sum_{s}{n_{s}s}}~.
        \label{S}
      \end{equation}
      Here, $n_{s}$ is the number of clusters of size $s$ per lattice site
      divided by the lattice volume,
      and the sums exclude the percolating cluster.
\end{itemize}

To determine the critical point of the percolation transition, we used 
the method suggested in \cite{Bin}.
For a given lattice size and 
a value of $\beta$  we counted how many times 
we found a percolating cluster. This number, divided
by the total number of configurations at that $\beta$,
is the  {\it percolation cumulant}.
This variable is a scaling function,
analogous to the Binder cumulant 
in continuous thermal phase transitions.
Curves corresponding to different lattice sizes
will thus cross at the critical point, apart from corrections to scaling.

The percolation cumulant gives not only the critical point $\beta_{c}$,
but also the critical exponent $\nu$.
In fact, if we rewrite the percolation cumulant as a function
of $(\beta-\beta_{c})L^{1/\nu}$, where $L$ is the linear spatial dimension
of the lattice, we should get a universal scaling curve
for all lattice sizes. We will adopt this method to determine
the percolation exponent $\nu$.

\section{Numerical Results}

As we are interested 
in the phase transition of $SU(2)$, we  first focus on the derivation of the
effective theory at the critical point.
Simulations were done on $32^3\times N_{\tau}$ lattices, with $N_{\tau}=2$
and $4$. 
For the critical coupling we took the value $\beta_c=1.8735(4)$
for $N_{\tau}=2$ as determined in \cite{aci}, and $\beta_c=2.29895(10)$
for $N_{\tau}=4$  from \cite{engels}.
The Monte Carlo update
we used consists in alternating one heat-bath and two overrelaxation steps.
We measured 
our observables every 60 updates, which makes
the analyzed configurations basically uncorrelated; the total number of
measurements was 2000.
Our aim is to check whether,
at $\beta=\beta_c$, we can find a projection
of the theory onto the spin model defined by Eq.(\ref{effham}).

For $N_{\tau}=2$ we considered two  sets of operator containing 10 and 15
spin-spin couplings, respectively.
The operators
connect a point ($000$) to ($100$), ($110$), ($111$),
($200$), ($210$), ($211$), ($220$),  ($221$), ($222$),
($300$), ($310$), ($311$), ($320$), ($321$), ($322$).
In the case of $N_{\tau}=4$ we have included further operators in our analysis,
which connect the point ($000$) to ($330$), ($331$), ($332$) and ($333$).
The final set of couplings is displayed in Table \ref{2couplings},
where I refers to the projection with 10 (15) operators for $N_{\tau}=2$
($N_{\tau}=4$) and II refers to the projection with 15 (19) operators for
$N_{\tau}=2$ ($N_{\tau}=4$). 
Note that the physical length scale changes with $N_{\tau}$. Two spins
separated by a distance $R$ measured in lattice units 
on the lattice with temporal extent $N_{\tau}$ are actually separated by 
distance $r$, with $r T_c = R/N_{\tau}$. 
Here $T_c=T(\beta_c)$
denotes the transition temperature of $SU(2)$ gauge theory.

\vskip0.5cm
\begin{table}[h]
  \begin{center}{
      \begin{tabular}{|c|c|c|c|c|}
\hline
\multicolumn{1}{|c|}{Couplings} &
\multicolumn{2}{|c|}{$N_{\tau}=2$} &
\multicolumn{2}{|c|}{$N_{\tau}=4$}  \\
\hline
~~~~~& I & II & I & II \\
\cline{2-5}
$\kappa_1$ (100) &~0.13042(8)~ &~0.13066(9)~  &~0.08385(3)~ &~0.08390(4)~\\
$\kappa_2$ (110) &~0.01911(1)~ &~0.01905(3)~  &~0.01842(4)~ &~0.01839(5)~\\
$\kappa_3$ (111) &~0.00476(5)~ &~0.00470(5)~  &~0.00769(5)~ &~0.00775(4)~\\
$\kappa_4$ (200) &~0.00794(5)~ &~~0.00801(11)~ &~0.00702(4)~ &~0.00697(1)~\\
$\kappa_5$ (210) &~0.00198(6)~ &~0.00192(4)~  &~0.00348(4)~ &~0.00343(2)~\\
$\kappa_6$ (211) &~0.00069(3)~ &~0.00062(8)~  &~0.00203(5)~ &~0.00197(1)~\\
$\kappa_7$ (220) &~0.00045(7)~ &~0.00033(2)~  &~0.00118(2)~ &~0.00114(1)~\\
$\kappa_8$ (221) &~0.00013(6)~ &~0.00007(2)~  &~0.00082(1)~ &~0.00083(1)~\\
$\kappa_9$ (222) &~0.00017(4)~ &~~0.00014(10)~ &~0.00038(7)~ &~0.00035(6)~\\
$\kappa_{10}$ (300) &~~0.00072(16)~ &~0.00058(3)~  &~0.00113(4)~ &~0.00105(9)~\\
$\kappa_{11}$ (310) &~~            &~0.00018(3)~  &~0.00076(5)~ &~0.00082(5)~\\
$\kappa_{12}$ (311) &~~            &~0.00008(1)~  &~0.00055(9)~ &~0.00055(4)~\\
$\kappa_{13}$ (320) &~~            &~0.00001(1)~  &~0.00032(3)~ &~0.00035(2)~\\
$\kappa_{14}$ (321) &~~            &~0.00006(1)~  &~0.00033(1)~ &~0.00030(4)~\\
$\kappa_{15}$ (322) &~~            &~-0.00005(6)~ &~0.00029(3)~ &~0.00013(4)~\\
$\kappa_{16}$ (330) &~~            &~~            &~~           &~0.00020(5)~\\
$\kappa_{17}$ (331) &~~            &~~            &~~           &~0.00018(3)~\\
$\kappa_{18}$ (332) &~~            &~~            &~~           &~0.00017(1)~\\
$\kappa_{19}$ (333) &~~            &~~            &~~           &~0.00017(4)~\\
\hline
      \end{tabular}
      }
\caption{Couplings of the effective theory
for the Ising-projected Polyakov loop configurations
of $(3+1)$-dimensional $SU(2)$. 
I and II denote the two sets of operator containing 10
and 15 operators for $N_{\tau}=2$, and 15 and 19 operators for $N_{\tau}=4$.}
\label{2couplings} 
  \end{center}
\end{table}

From Table 1 one can see that the difference between couplings
calculated for set I and set II is of the order of the statistical 
errors. 
This implies that possible truncation errors are small.
Comparing our couplings for set I and $N_{\tau}=2$ with the
corresponding couplings obtained in \cite{okawa88}, we see that they agree
within statistical errors, although multi-spin couplings
were also included in the analysis of Ref. \cite{okawa88}. 
This suggests that 
the multi-spin couplings are not important, for $N_{\tau}=2$;
in fact, most of them are compatible with zero within statistical errors.
Note that because of much improved statistics, 
the errors on the couplings in Table 1 
are by an order of magnitude smaller than in 
\cite{okawa88}. 
Since the error on $\kappa_{15}$ is of the order of its
average value, we can set $\kappa_{15}=0$ 
without appreciable effects. In this way, we indeed have obtained 
the effective theory we were looking for, with
only ferromagnetic spin-spin interactions.

\begin{figure}[h]
\begin{center}
\epsfig{file=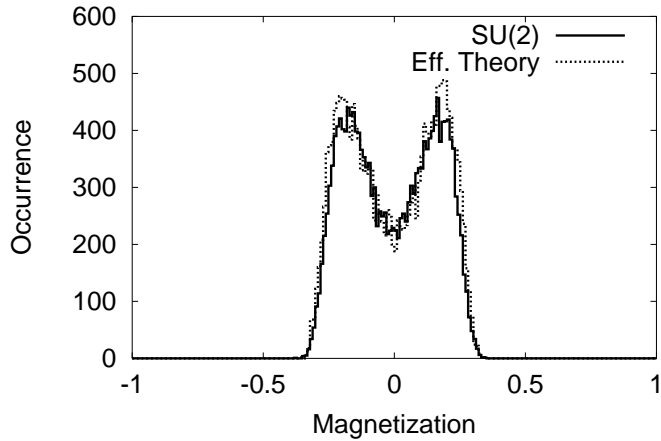,  width=9cm}
\caption[Comparison of the magnetization histograms derived
by the Polyakov loop configurations and by the effective
theory: $N_{\tau}=2$]
{Comparison of the magnetization histograms derived
from the Polyakov loop configurations and from the effective
theory (\ref{effham}) defined by the couplings of Table \ref{2couplings}.}
\label{histo}
\end{center}
\end{figure}

Fig. \ref{histo} shows a comparison between the 
magnetization distribution for $N_{\tau}=2$ of the Polyakov loop configurations
and the one of the effective theory containing 15 operators:
the two histograms are very similar. The values  
of the average magnetization $m$ 
are also in agreement: for $SU(2)$, $m=0.091(1)$ and 
for the spin model, $m=0.0923(7)$. The average magnetization $m$ calculated
from the effective theory containing 10 operators is by $10 \%$ smaller than
this value. One can also clearly see differences between the magnetization
distributions calculated from the Polyakov loop configurations and from the
effective theory with 10 operators.

In the case of $N_{\tau}=4$ we have considered two sets of spin-spin operators
containing 15 and 19 couplings labeled I and II, respectively. 
The values of the couplings are
shown in Table  \ref{2couplings}.  
Note that all operators are positive
in this case. 
The average value of the magnetization $m$ obtained from the effective
theory with 19 operators is $0.121(3)$, which should be compared with the
corresponding value calculated from SU(2) configurations, $m=0.128(6)$.   
The effective theory with 15 operators gives a value which is by $20 \%$ smaller.

The couplings reported in Table 1 decrease exponentially
as function of $r T$. The exponential fall-off is approximately
described by $\sim \exp(-9 rT)$ for both $N_{\tau}=2$ and $N_{\tau}=4$.
It is easy to see that couplings of the effective spin 
theory at fixed $r T$ should
scale as $N_{\tau}^{-4}$ close to the continuum limit. A comparison of 
couplings for $N_{\tau}=2$ and $N_{\tau}=4$ at the same value of $r T_c$ shows
that this scaling holds up to $20 \%$ scaling violation. 
This observation suggests that the form of the effective theory
for obtained $N_{\tau}=2$ and $4$ will not change qualitatively 
by approaching the continuum limit. In particular, we expect that the effective
theory will contain only ferromagnetic spin-spin coupling 
in the continuum limit.

To check the volume dependence of the couplings we also determined them 
on $16^3\times 2$ and $16^3 \times 4$ lattices. 
The resulting values  agree with those obtained
from simulations on larger $32^3 \times2$ and $32^3 \times 4$ lattices within
statistical errors. We therefore conclude that the set of couplings
we calculated is, with good approximation, also that  
of the effective theory in the infinite volume limit,
which is what we will need for our subsequent percolation studies
\footnote{The bond weights are 
determined by the temperature. 
Once we fix the temperature, for different
lattice sizes we must use the same set of 
values for the bond weights.}.

We have shown that the effective $Z(2)$ theory with 15 ($N_{\tau}=2$)
and 19  ($N_{\tau}=4$) spin-spin operators gives a good description
of the thermal critical behaviour of $SU(2)$ gauge theory. We now 
want to see if this effective theory also allows us to extend the Polyakov
loop percolation description of the deconfinement transition 
\cite{satz98,fortunato00} from the strong coupling limit \cite{fortunato00}
to a more general situation closer to the continuum limit.
To study this we generate $SU(2)$ Polyakov loop configuration on 
$N_{\sigma}^3 \times N_{\tau}$ lattices, reduce them to $Z(2)$ spin 
configurations and then define clusters with the bond weights as defined 
in Eq. (\ref{tu}). A cluster 
thus consists of a set of aligned spins coupled by one or more
bond weights. We want to determine the percolation behaviour of these
clusters. 

We stress that the bond weights are 
temperature-dependent. Our 
effective theory represents a projection
of $SU(2)$ for $\beta=\beta_c$. But
in order to carry out
our analysis, we need to evaluate the 
percolation variables at different values of $\beta$.
Strictly speaking, for each $\beta_{i}$
we should derive the corresponding 
effective theory and use the 
set $\{\kappa(\beta_i)\}$ to 
calculate the bond weights (\ref{tu}) at $\beta_{i}$. 
We have calculated the couplings of the effective theory for several
values of $\beta$ in the interval [1.873,1.883] for $16^3 \times 2$ lattice.
It turned out
that the variation of the couplings due to the variation of 
$\beta$ in this small interval
is of the order of the statistical errors.
Because of that, at each $\beta$,
we shall use the same set of bond probabilities,
determined by the couplings
of Table \ref{2couplings}.

\begin{figure}
\centerline{\epsfig{file=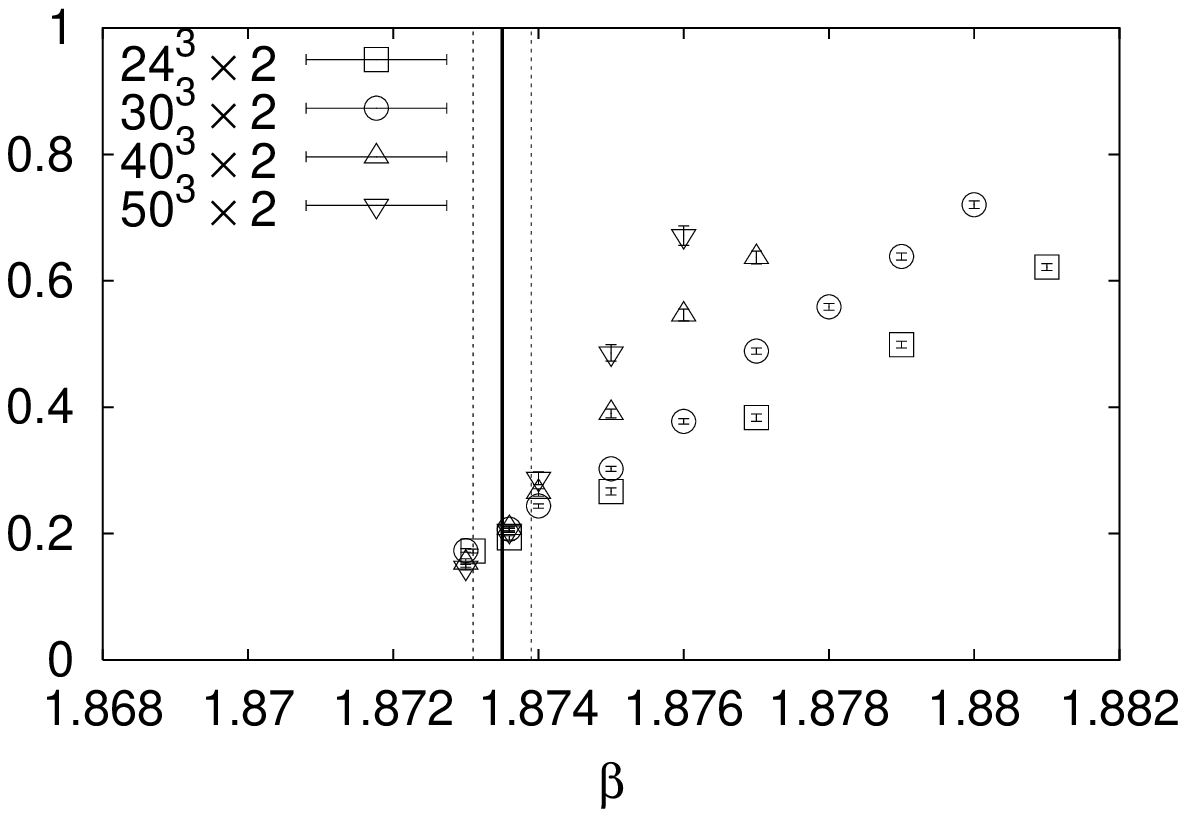,width=8cm} \epsfig{file=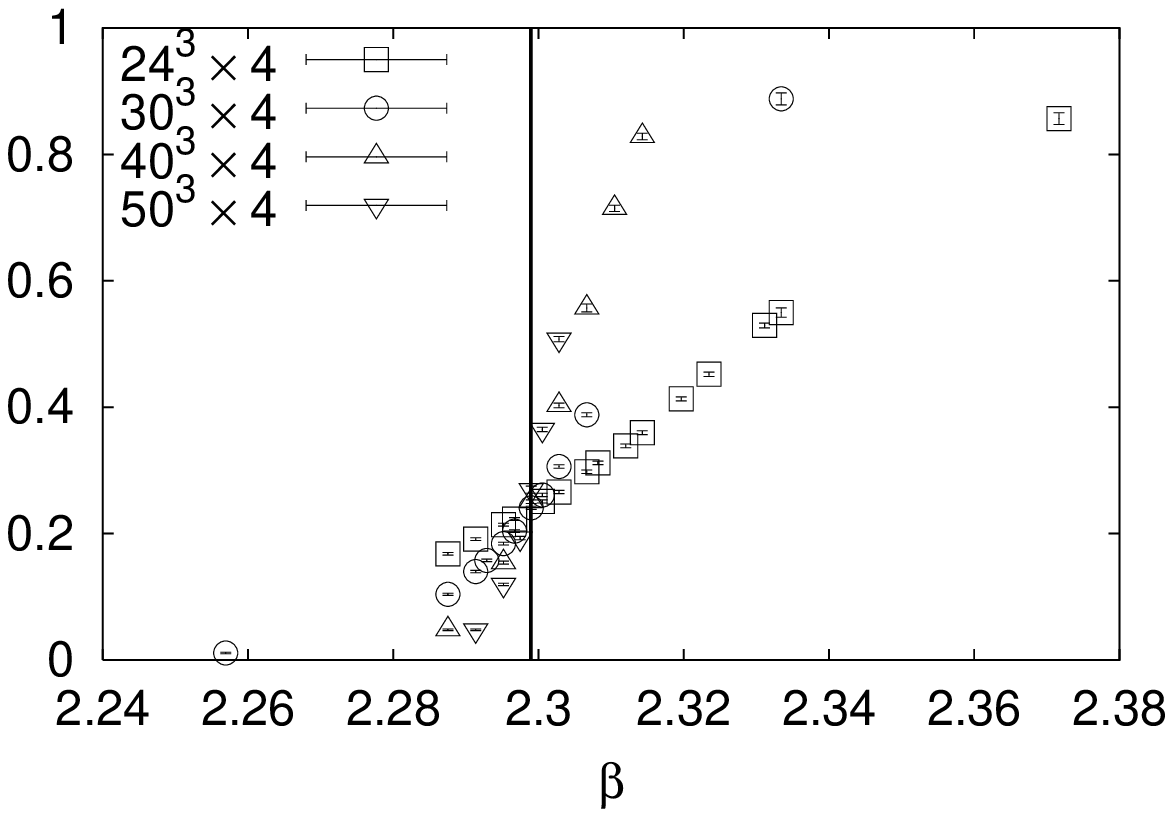,width=8cm}} 
\vskip-0.2cm
\caption{ Percolation cumulants near 
the critical point for different spatial volume for $N_{\tau}=2$ (left) and
for $N_{\tau}=4$ (right). The solid vertical lines indicate the value of the
thermal threshold. For $N_{\tau}=2$ we also indicate the uncertainty on the 
threshold by two dashed lines. For $N_{\tau}=4$ the errors on the critical
coupling are too small to be visible in the Figure.}
\vspace*{-0.5truecm}
\label{sceffnt2}
\end{figure}

For the analysis of the percolation transition we have used four different 
lattices of size $N_{\sigma}^3\times N_{\tau}$, with $N_{\sigma}=24,30,40,50$
and  $N_{\tau}=2,4$.
Fig. \ref{sceffnt2} shows
the behavior of the percolation cumulant
as a function of $\beta$. In both cases the curves cross remarkably well at
the same point,
in excellent agreement with the thermal thresholds, indicated
by the solid line.

Next, we performed further simulations near criticality in order to study
the critical exponents and to
determine more precisely the value  of the geometrical threshold.
From standard finite size scaling fits
of the average cluster size $S$ at different $\beta$ values
we determine the value $\beta_c$ at which 
the scaling fit gives the best $\chi^2$; the slope 
of the straight line of the log-log plot of $S$ at $\beta_c$ gives $\gamma/\nu$. 
The errors on $\beta_c$ and on the
ratio of critical exponents are calculated by determining
the $\beta$-range containing $\beta_c$ such that for each $\beta$
one still gets a good $\chi^2$ for the scaling fit.
Unfortunately
we could not determine $\beta/\nu$, because of strong 
fluctuations of the percolation strength $P$ around $\beta_c$.
The value of $P$ at criticality is in general quite small, and it
is more strongly influenced by the approximations involved in
our approach. Furthermore, we determine the exponent $\nu$
from the finite size scaling of the pseudo-critical points,
i. e. the positions of the peak of $S$ for the different lattices we considered.

\begin{figure}[h]
\centerline{\epsfig{file=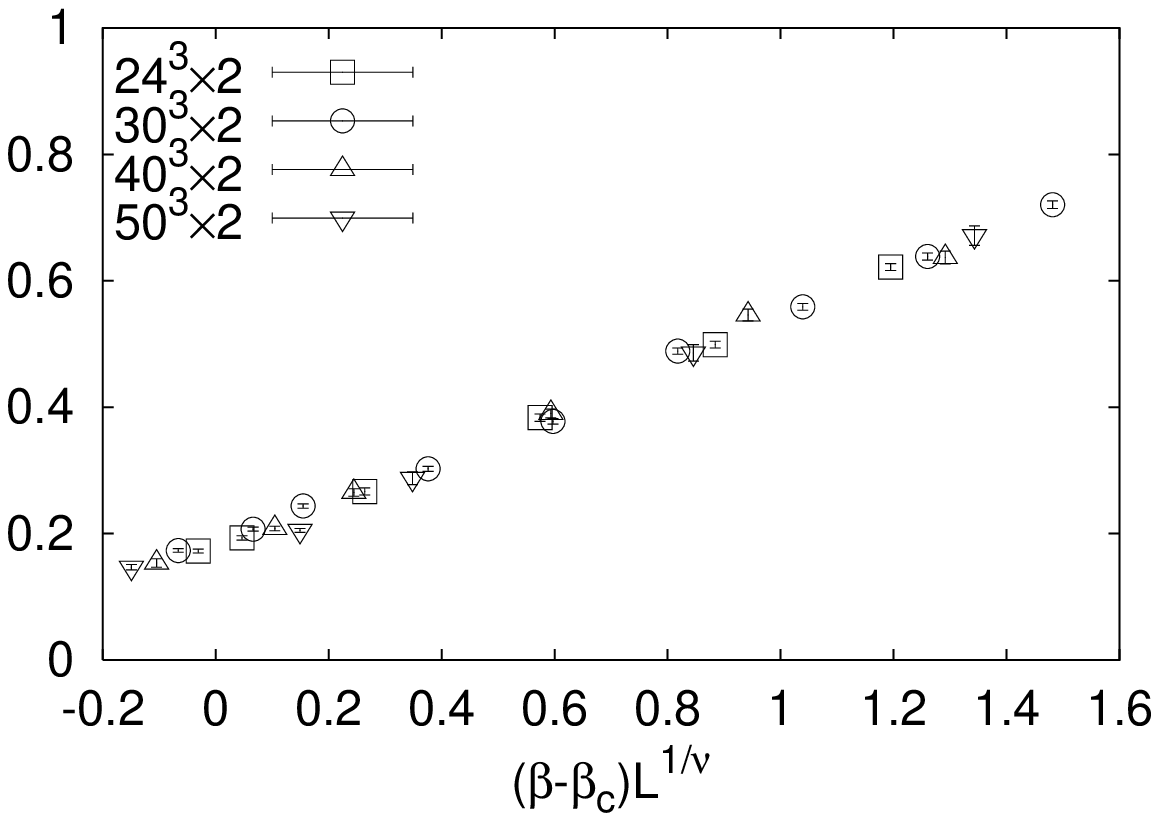,width=8cm}, \epsfig{file=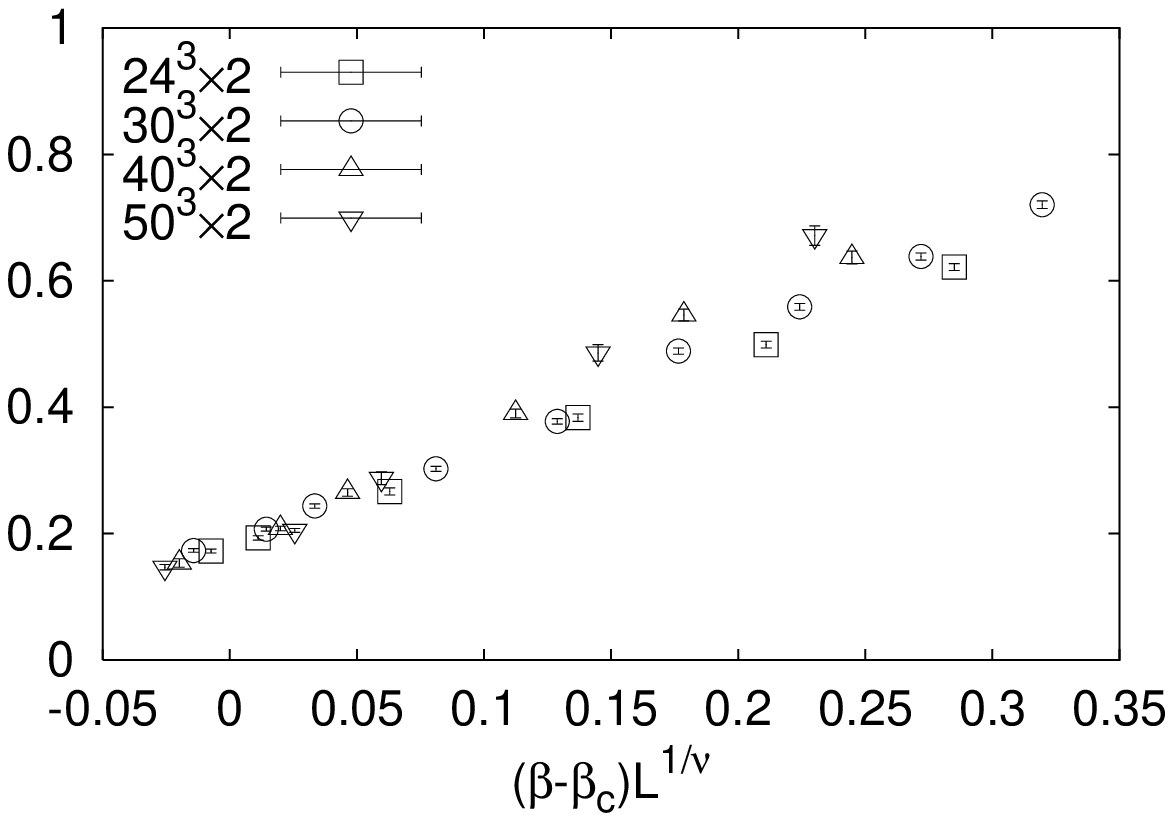,  width=8cm}}
\caption{ Rescaling of the percolation cumulant 
curves shown in Fig. \ref{sceffnt2} for $N_{\tau}=2$ with 
the 3-dimensional Ising exponent $\nu=0.6294$ (left) and with 
the 3-dimensional random percolation exponent $\nu=0.88$ (right),
using $\beta_{c}=1.8734$ for the critical coupling.}
\label{rescnt2}
\end{figure}

\begin{figure}[h]
\centerline{\epsfig{file=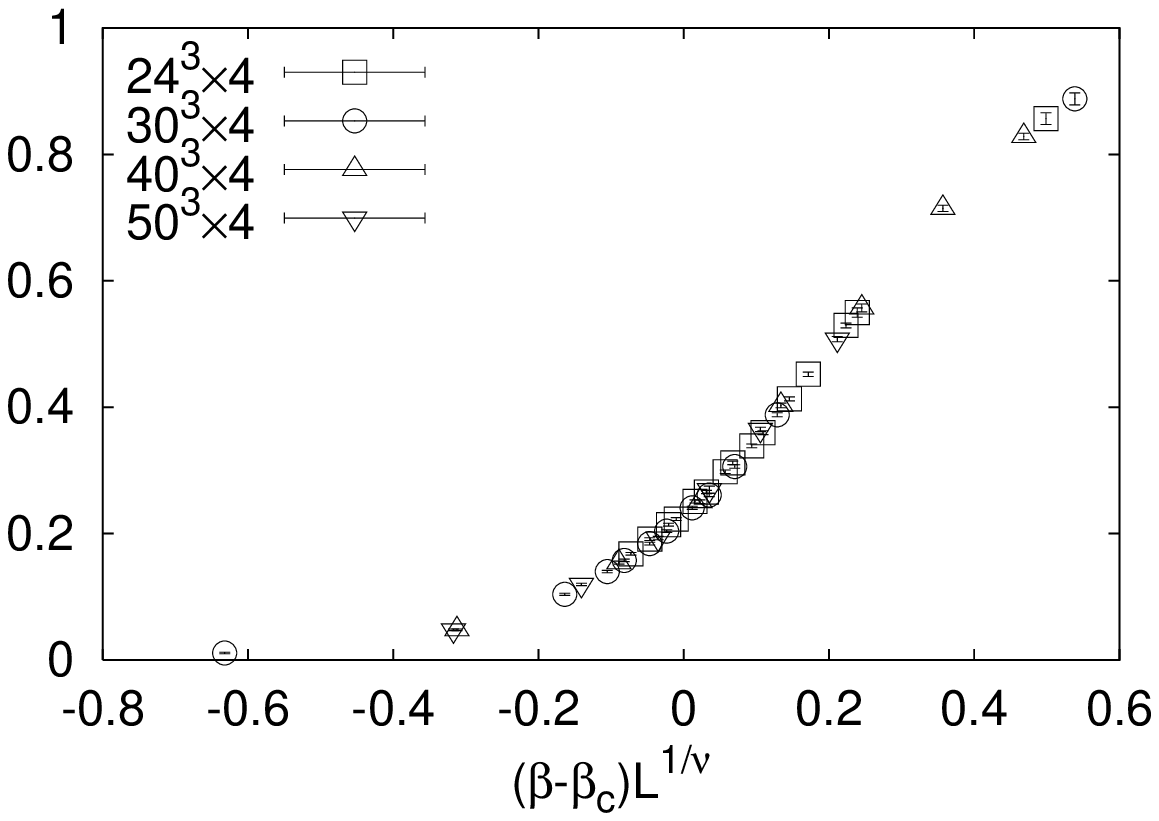,width=8cm}, \epsfig{file=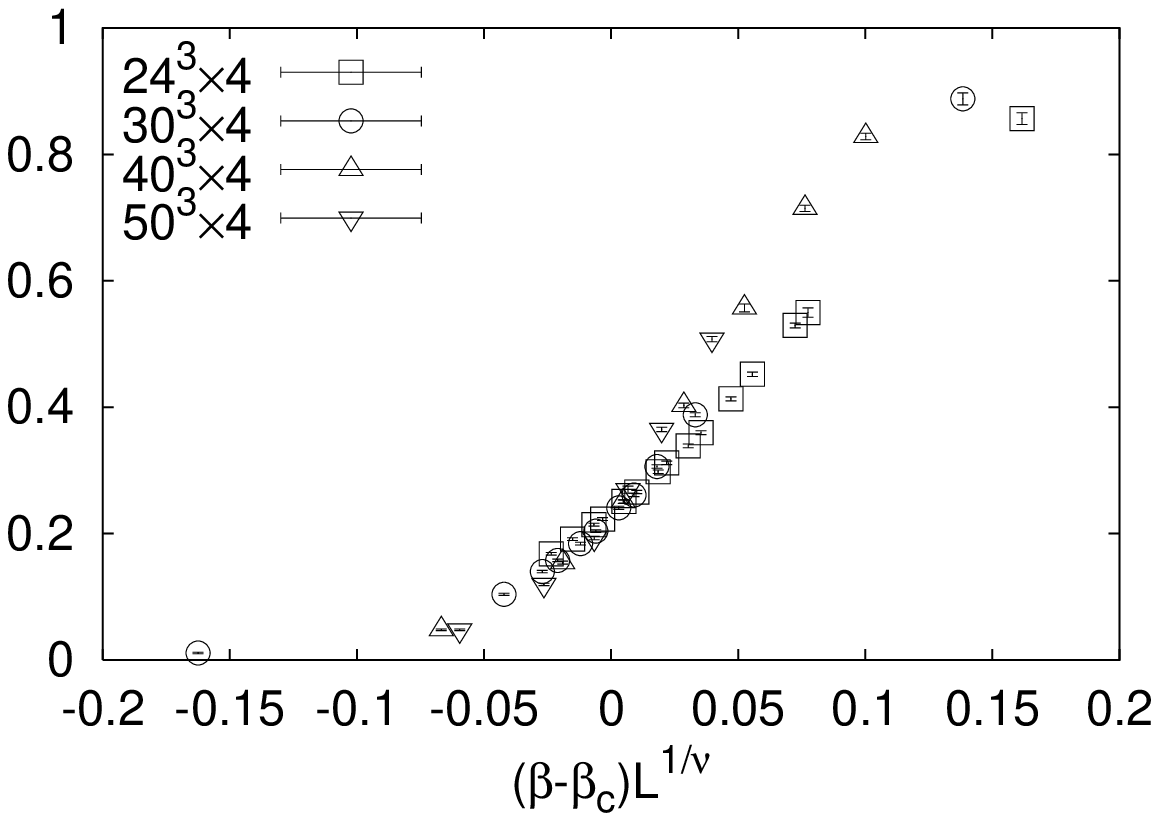,  width=8cm}}
\caption{ Rescaling of the percolation cumulant 
curves shown in Fig. \ref{sceffnt2} for $N_{\tau}=4$ with 
the 3-dimensional Ising exponent $\nu=0.6294$ (left) and with 
the 3-dimensional random percolation exponent $\nu=0.88$ (right),
using $\beta_{c}=2.2991$ for the critical coupling.}
\label{rescnt4}
\end{figure}

The percolation threshold and the critical exponents are summarized
in Table \ref{tabeffnt2} for both $N_{\tau}=2$ and $4$.
There we also give the critical exponents of the thermal $SU(2)$ transition and
those corresponding to the 3D Ising model. As one can see from Table 2, 
good agreement between the percolation exponents and the 3D Ising 
exponents is
found. In Fig. \ref{rescnt2} we show the scaling of the percolation
cumulant with the Ising exponent and with the 3-dimensional random percolation exponent
$\nu=0.88$ for the case of $N_{\tau}=2$.
The figure shows that scaling is quite consistent with the Ising
exponent while the random percolation exponent is ruled out by the data.
The situation is the same for $N_{\tau}=4$ as one can see in Fig. \ref{rescnt4}.

\begin{table}[h]
  \begin{center}{
      \begin{tabular}{|c|c|c|c|c|c|c|}
\hline
\multicolumn{1}{|c|}{~~} &
\multicolumn{2}{|c|}{Critical point} & 
\multicolumn{2}{|c|}{${\gamma}/{\nu}$} & 
\multicolumn{2}{|c|}{$\nu$}\\ 
\cline{2-7}
~~& $N_{\tau}=2$ & $N_{\tau}=4$ & $N_{\tau}=2$ & $N_{\tau}=4$ & $N_{\tau}=2$ &
$N_{\tau}=4$  \\
\hline
$\vphantom{\displaystyle\frac{1}{1}}$  Percolation&
$1.8734(2)$ & $2.2991(2)$ & $1.977(17)$ & $1.979(16)$ & 
$0.628(11)$ & $0.629^(0.011)$\\   
\hline
$\vphantom{\displaystyle\frac{1}{1}}$ Thermal& 
$1.8735(4)$ & $2.29895(10)$ & $1.953(9)$ & $1.944(13)$ & 
$0.630^(10)$ &$0.630(11)$ \\ 
\hline
\multicolumn{1}{|c|}{3D Ising} & 
\multicolumn{2}{|c|}{} & 
\multicolumn{2}{|c|}{1.967(7)}&
\multicolumn{2}{|c|}{0.6294(10)}\\ 
\hline
      \end{tabular}
      }
\caption {\label{tabeffnt2} 
Percolation critical indices for 
$(3+1)$-dimensional $SU(2)$.
For comparison we also list the thermal results \cite{aci}
and the 3D Ising values \cite{parisi}.} 
  \end{center}
\end{table}

\section{Conclusions}
We have shown that in the vicinity of the critical point 
$SU(2)$ gauge theory can be well approximated by a simple $Z(2)$ model
with ferromagnetic spin-spin interactions alone. We have tested 
the accuracy of the projection for 
$N_{\tau}=2$ and $4$. In both cases we have found
that there is basically no trace of frustration.
The percolation picture of the effective model 
is suitable to describe the deconfining transition of
$SU(2)$ in purely geometrical terms.

In principle the present approach can be extended also to larger
values of  $N_{\tau}$ (smaller lattice spacing). We expect that 
for larger $N_{\tau}$  more operators should be included in
the effective action. This can be avoided by first performing a block
spin transformation of the Polyakov loop configurations and then
doing the whole analysis for the blocked configurations. 
We have shown that the couplings of the effective theory decrease
exponentially in strength as function of $r T_c$ and their relative
magnitude scales with $N_{\tau}^{-4}$. This suggests that the effective
theory remains well defined in the continuum limit and will stay
ferromagnetic and short ranged.
Therefore we do
not expect that the present picture will change 
qualitatively for larger $N_{\tau}$.

\section*{Ackowledgements}

The work has been supported by the TMR network ERBFMRX-CT-970122
and the DFG under grant Ka 1198/4-1.


\end{document}